\documentclass[11pt,a4paper]{article}
\usepackage[hyperref]{emnlp2018}
\usepackage{times}
\usepackage{latexsym}
\usepackage{arydshln}
\usepackage{url}
\usepackage{graphicx}
\usepackage{comment}
\usepackage{xspace}

\aclfinalcopy 

\newcommand{\st}[1]{\ {\tiny \mbox{(#1)}}\xspace}
\newcommand{\mlp}{\textsc{mlp}\xspace}
\newcommand{\lstm}{\textsc{lstm}\xspace}
\newcommand{\bilstm}{\textsc{bilstm}\xspace}
\newcommand{\rnn}{\textsc{rnn}\xspace}
\newcommand{\cnn}{\textsc{cnn}\xspace}

\newcommand{\ir}{\textsc{ir}\xspace}
\newcommand{\nlp}{\textsc{nlp}\xspace}

\newcommand{\bioasq}{\textsc{bioasq}\xspace}
\newcommand{\trec}{\textsc{trec}\xspace}
\newcommand{\trecrob}{\textsc{trec robust}\xspace}
\newcommand{\drmm}{\textsc{drmm}\xspace}
\newcommand{\pacrr}{\textsc{pacrr}\xspace}
\newcommand{\pacrrdrmm}{\textsc{pacrr-drmm}\xspace}
\newcommand{\abeldrmm}{\textsc{abel-drmm}\xspace}
\newcommand{\abeldrmmmv}{\textsc{abel-drmm+mv}\xspace}
\newcommand{\positdrmm}{\textsc{posit-drmm}\xspace}
\newcommand{\positdrmmmv}{\textsc{posit-drmm+mv}\xspace}
\newcommand{\bmtf}{\textsc{bm25}\xspace}
\newcommand{\idf}{\textsc{idf}\xspace}
\newcommand{\map}{\textsc{map}\xspace}

\newcommand{\medline}{\textsc{medline}\xspace}

\title{Deep Relevance Ranking Using Enhanced Document-Query Interactions}

\author{Ryan McDonald$^{1,2}$\textnormal{,} 
Georgios-Ioannis Brokos$^1$ \and Ion Androutsopoulos$^1$\\
\\
$^1$Dept.\ of Informatics, Athens University of Economics and Business, Greece\\
$^2$Google AI\\
}
\date{}

\begin{document}
\maketitle

\begin{abstract}
We explore several new models for document relevance ranking, 
building upon the Deep Relevance Matching Model (\drmm) of \newcite{guo2016deep}. Unlike \drmm, which uses  
context-insensitive encodings of terms and query-document term interactions, we inject rich context-sensitive encodings throughout our models, inspired by \pacrr's \cite{hui2017pacrr} convolutional $n$-gram matching features, but extended in several ways including  multiple views of query and document inputs.  We test our models on datasets from the \bioasq question answering challenge \cite{tsatsaronis2015overview} and \trecrob 2004 \cite{voorhees2005trec}, showing they outperform \bmtf-based
baselines, \drmm, and \pacrr.\\
\end{abstract}


\section{Introduction}

Document relevance ranking, also known as \emph{ad-hoc retrieval} \cite{harman2005trec}, is the task of ranking documents from a large collection using the query and the text of 
each document only. This contrasts with standard information retrieval (\ir) systems that rely on text-based signals in conjunction with network structure \cite{page1999pagerank,kleinberg1999authoritative} and/or user feedback \cite{,joachims2002optimizing}.  
Text-based ranking is particularly important when (i) click-logs do not exist or are small, and (ii) the network structure of the collection is 
non-existent or not informative for query-focused relevance. Examples include various domains in digital libraries, e.g., patents \cite{azzopardi2010search} or scientific literature \cite{wu2015citeseerx,tsatsaronis2015overview}; enterprise search \cite{hawking2004challenges}; and personal search \cite{chirita2005activity}.

\begin{figure}[t] 
\includegraphics[width=3in]{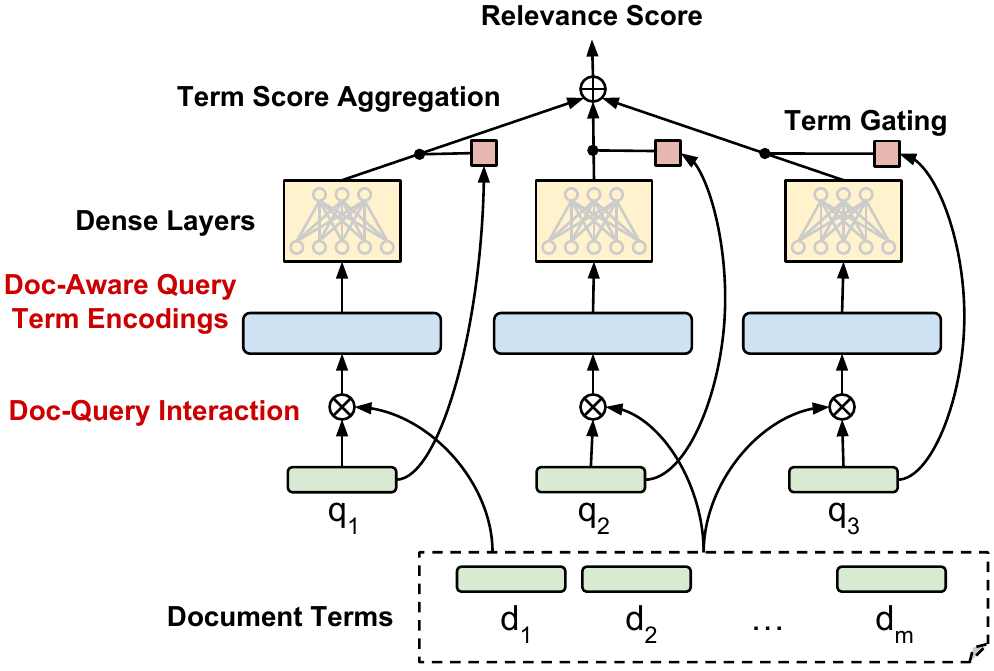}
\vspace{-0.15in}
\caption{Illustration of \drmm \cite{guo2016deep} for a query of three terms and a document of $m$ terms.}
\vspace{-4mm}
\label{fig:top-level-drmm}
\end{figure}

We investigate new deep learning architectures for document relevance ranking, focusing on 
\emph{term-based interaction models}, where
query terms (\emph{q-terms} for brevity) are scored relative to a document's terms (\emph{d-terms}) and their scores are aggregated to produce a relevance score for the document. Specifically, we use the Deep Relevance Matching Model (\drmm) of \newcite{guo2016deep} (Fig.~\ref{fig:top-level-drmm}), which was shown to outperform  
strong \ir baselines and other recent deep learning methods. 
\drmm uses pre-trained word embeddings for 
q-terms and d-terms, and cosine similarity histograms (outputs of $\otimes$ in Fig.~\ref{fig:top-level-drmm}), each capturing the similarity of a 
q-term to all the d-terms of a particular document. 
The histograms are fed to an \mlp
(dense layers of Fig.~\ref{fig:top-level-drmm})
that produces the (document-aware) score of each q-term.
Each q-term score is then weighted using a gating mechanism (topmost box nodes in Fig.~\ref{fig:top-level-drmm}) that examines properties of the q-term to assess its importance for ranking (e.g., common words are less important). The sum of the weighted q-term scores is the relevance score of the document.
This ignores entirely the contexts where the terms occur, in contrast to recent \emph{position-aware} models such as \pacrr \cite{hui2017pacrr} or those based on recurrent representations \cite{palangi2016deep}.

In order to enrich \drmm with context-sensitive representations, we need to change fundamentally how q-terms are scored. This is because rich context-sensitive representations -- such as input term encodings based on \rnn{s} or \cnn{s} -- require end-to-end training and histogram construction is not differentiable. To account for this we investigate novel query-document interaction mechanisms that are differentiable and show empirically that they are effective ways to enable end-to-end training of context-sensitive \drmm models. This is the primary contribution of this paper.

Overall, we explore several extensions to \drmm, including: \pacrr-like convolutional $n$-gram matching
features ($\S$\ref{sec:pacrr-drmm});
context-sensitive term encodings ($\S$\ref{sec:input-encodings}); 
query-focused attention-based document representations ($\S$\ref{sec:had-drmm});
pooling to reward denser term matches and turn rich term representations into fixed-width vectors ($\S$\ref{sec:ps-drmm});
multiple views of terms, e.g., context sensitive, insensitive, exact matches ($\S$\ref{sec:multiview}).

We test our models on data from the \bioasq biomedical question answering challenge \cite{tsatsaronis2015overview} 
and \trecrob 2004 \cite{voorhees2005trec}, showing that they outperform strong \bmtf-based baselines \cite{robertson2009probabilistic}, \drmm, and \pacrr.\footnote{The code and data of our experiments, including word embeddings, are available at \url{https://github.com/nlpaueb/deep-relevance-ranking}.}


\section{Related Work}
\label{sec:related}

Document ranking has been studied since the dawn of \ir; classic term-weighting schemes were designed for this problem \cite{sparck1972statistical,robertson1976relevance}. With the advent of statistical \nlp and statistical \ir, probabilistic language and topic modeling were explored \cite{zhai2001study,wei2006lda}, followed  recently by deep learning \ir methods \cite{lu2013deep,hu2014convolutional,palangi2016deep,guo2016deep,hui2017pacrr}.

Most document relevance ranking methods fall within two categories: representation-based, e.g., \newcite{palangi2016deep}, or interaction-based, e.g., \newcite{lu2013deep}. In the former, representations of the query and document are generated independently. Interaction between the two only happens at the final stage, where a score is generated indicating relevance. End-to-end learning and backpropagation through the network tie the two representations together. In the interaction-based paradigm, 
explicit encodings between pairs of queries and documents are induced. This allows direct modeling of exact- or near-matching terms (e.g., synonyms), which is crucial for relevance ranking.
Indeed, \newcite{guo2016deep} showed that the interaction-based \drmm  outperforms previous representation-based methods. 
On the other hand, interaction-based models are less efficient, since one cannot index a document representation independently of the query. This is less important, though, when relevance ranking methods rerank the top documents returned by a conventional \ir engine, which is the scenario we consider here.

One set of our experiments ranks biomedical texts. Several methods have been proposed for the \bioasq challenge \cite{tsatsaronis2015overview}, mostly based on traditional \ir techniques. The most related work is of \newcite{mohan2017deep}, who use a deep learning architecture. Unlike our work, they focus on 
user click data as a supervised signal, and they use 
context-insensitive representations of document-query interactions.
The other dataset we experiment with, \trecrob 2004 \cite{voorhees2005trec}, has been used extensively to evaluate 
traditional and deep learning \ir methods. 

Document relevance ranking is also related to other \nlp tasks. Passage scoring for question answering \cite{surdeanu2008learning} ranks passages 
by their relevance to the question; several deep networks have been proposed, e.g., \newcite{tan2015lstm}.
Short-text matching/ranking is also related and has seen recent deep learning solutions \cite{lu2013deep,hu2014convolutional,severyn2015learning}.
In document relevance ranking, though, documents are typically much longer than queries, which makes methods from other tasks that consider pairs of short texts not directly applicable.

Our starting point is \drmm, to which we add richer representations inspired by \pacrr. Hence, we first discuss \drmm and \pacrr further. 


\subsection{DRMM}
\label{sec:drmm}

We have already presented an overview of \drmm. For gating (topmost box nodes of Fig.~\ref{fig:top-level-drmm}), \newcite{guo2016deep} use a linear self-attention:
\[g_i = \textrm{softmax}\Big(w_g^T \, \phi_g(q_i); q_1, \dots, q_n\Big)\]
where $\phi_g(q_i)$ is the embedding $e(q_i)$ of the $i$-th q-term, 
or its \idf, $\textrm{idf}(q_i)$, and $w_g$ is a weights vector. Gating aims to weight the (document-aware) score of each q-term
(outputs of dense layers in Fig.~\ref{fig:top-level-drmm}) based on the importance of the term. We found that   
$\phi_g(q_i) = [e(q_i); \textrm{idf}(q_i)]$, where `;' is concatenation, was optimal for all \drmm-based models.

The crux of the original \drmm are the bucketed cosine similarity histograms (outputs of $\otimes$ nodes in Fig.~\ref{fig:top-level-drmm}), each capturing the similarity of a 
q-term to all the d-terms. In each histogram, each bucket counts the number of d-terms whose cosine similarity to the q-term is within a particular range. Consider a document with three terms, with cosine similarities, \textit{s}, to a particular q-term $q_i$ $0.5$, $0.1$, $-0.3$, respectively. If we used two buckets 
$-1 \leq s < 0$ and $0 \leq s \leq 1$,
then the input to the dense layers for $q_i$ would be $\langle 1, 2 \rangle$.
The fixed number of buckets leads to a fixed-dimension input for the dense layers and makes the model agnostic to different document and query lengths -- one of \drmm's main strengths.
The main disadvantage is that bucketed similarities are independent of the contexts where terms occur. 
A q-term `regulated' will have a perfect match with a d-term `regulated', even if the former is `up regulated' and the latter is `down regulated' in context. 
Also, there is no reward for term matches that preserve word order, or multiple matches within a small window.


\subsection{PACRR}
\label{sec:pacrr}

\begin{figure}[t]
\includegraphics[width=3in]{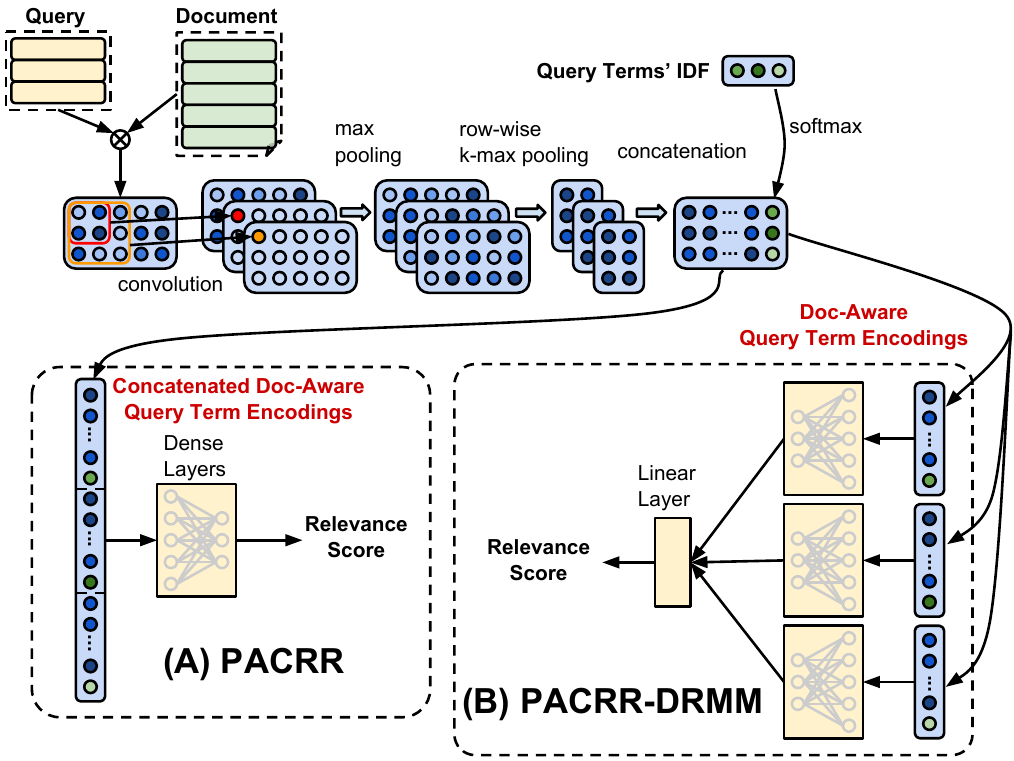}
\vspace{-0.17in}
\caption{\pacrr \cite{hui2017pacrr} and \pacrrdrmm.}
\vspace{-4mm}
\label{fig:pacrr}
\end{figure}

In \pacrr \cite{hui2017pacrr}, a query-document term similarity matrix 
$\textit{sim}$ is computed (Fig.~\ref{fig:pacrr}A). Each cell $(i, j)$ of 
$\textit{sim}$ contains the cosine similarity between the embeddings of a
q-term $q_i$ and a d-term $d_j$. 
To keep the dimensions $l_q \times l_d$ of 
$\textit{sim}$ fixed across queries and documents of varying lengths, queries are padded to the maximum number of 
q-terms $l_q$, and only 
the first $l_d$ terms per document are retained.\footnote{We use \textsc{pacrr}-\textit{firstk}, which 
\newcite{hui2017pacrr} recommend when documents fit in memory, as in our experiments.}
Then, convolutions (Fig.~\ref{fig:pacrr}A) of different kernel sizes $n\times n$ ($n=2,\dots,l_g$) are applied to 
$\textit{sim}$ to capture $n$-gram query-document similarities. For each size  $n\times n$, multiple kernels (filters) are used. Max pooling is then applied along the dimension of the filters (max value of all filters), 
followed by row-wise $k$-max pooling to capture the strongest $k$ signals between each 
q-term and all the d-terms. 
The resulting matrices are concatenated into a single matrix where each row is a 
document-aware q-term encoding; the \idf of the q-term is also appended, normalized by applying a softmax across the \idf{s} of all the q-terms.
Following \newcite{hui2018copacrr}, we concatenate the rows of the resulting matrix into a single vector, which is passed to an \mlp (Fig.~\ref{fig:pacrr}A, dense layers) that produces a query-document relevance score.\footnote{\newcite{hui2017pacrr} used an additional \lstm, which was later replaced by the final concatanation \cite{hui2018copacrr}.}

The primary advantage of \pacrr over \drmm is that it models context via the $n$-gram convolutions, i.e., denser $n$-gram matches and matches preserving word order are encoded. However, this context-sensitivity is weak, as the convolutions operate over the similarity matrix, not directly on terms or even term embeddings. Also, unlike \drmm, \pacrr requires padding and hyperparameters for maximum number of q-terms ($l_q$) and d-terms ($l_d)$, since 
the convolutional and dense layers operate over fixed-size matrices and vectors.
On the other hand, \pacrr is end-to-end trainable -- though \newcite{hui2017pacrr} use fixed pre-trained embeddings -- unlike \drmm where the bucketed histograms are not differentiable. 


\section{New Relevance Ranking Models}


\subsection{PACRR-DRMM}
\label{sec:pacrr-drmm}

In a \drmm-like version of \pacrr, instead of 
using an \mlp (dense layers, Fig.~\ref{fig:pacrr}A) to score the 
\emph{concatenation} of all the (document-aware) q-term encodings, the \mlp \emph{independently} scores each q-term encoding
(the same \mlp for all q-terms, Fig.~\ref{fig:pacrr}B); the resulting scores are aggregated via a linear layer. This version, \pacrrdrmm, performs better than \pacrr, using the same number of hidden layers in the \mlp{s}. Likely this is due to the fewer parameters of its \mlp, which is shared across the q-term representations and operates on shorter input vectors. Indeed, in early experiments \pacrrdrmm was less prone to over-fitting. 

In \pacrrdrmm, the scores of the q-terms (outputs of dense layers, Fig.~\ref{fig:pacrr}B) are not weighted by a gating mechanism, unlike \drmm (Fig.~\ref{fig:top-level-drmm}). Nevertheless, the \idf{s} of the q-terms, which are appended to the q-term encodings (Fig.~\ref{fig:pacrr}B), are a form of term-gating (shortcut passing on information about the terms, here their \idf{s}, to upper layers) applied before scoring the q-terms. By contrast, in \drmm (Fig.~\ref{fig:top-level-drmm}) term-gating is applied after q-term scoring, and operates on $[e(q_i); \textrm{idf}(q_i)]$.


\subsection{Context-sensitive Term Encodings}
\label{sec:input-encodings}

In their original incarnations, \drmm and \pacrr use pre-trained word embeddings that are insensitive to the context of a particular query or document where a term occurs. This contrasts with the plethora of systems that use context-sensitive word encodings (for each particular occurrence of a word) in virtually all \nlp tasks \cite{bahdanau2014neural,plank2016multilingual,lample2016neural}. In general, this is achieved via \rnn{s}, e.g., \lstm{s} \cite{gers1999learning}, or \cnn{s} \cite{bai2018empirical}.

In the \ir literature, context-sensitivity is typically viewed through two lenses: term proximity \cite{buttcher2006term} and term dependency \cite{metzler2005markov}. The former assumes that
the context around a term match is also relevant, whereas the latter aims to capture when multiple terms (e.g., an $n$-gram) must be 
matched together. An advantage of neural network architectures like \rnn{s} and \cnn{s} is that they can capture both.

In the models below ($\S\S$\ref{sec:had-drmm}--\ref{sec:ps-drmm}), an encoder produces the \emph{context-sensitive encoding} of each q-term or d-term from the pre-trained embeddings. To compute this we use a standard \bilstm encoding scheme and set the context-sentence encoding as the concatenation of the last layer's hidden states of the forward and backward \lstm{s} at each position. As is common for \cnn{s} and even recent \rnn term encodings \cite{peters2018deep}, we use the original term embedding $e(t_i)$ as a residual and combine it with the \bilstm encodings. Specifically, if $\overrightarrow{h}(t_i)$ and $\overleftarrow{h}(t_i)$ are the last layer's hidden states of the left-to-right and right-to-left \lstm{s} for term $t_i$, respectively, then we set the context-sensitive term encoding as:
\begin{equation}
c(t_i) = [\overrightarrow{h}(t_i) + e(t_i); \overleftarrow{h}(t_i) + e(t_i)]
\label{eq:ct}
\end{equation}
Since we are adding the original term embedding to each \lstm hidden state, we require the dimensionality of the hidden layers to be equal to that of the original embedding. Other methods were tried, including passing all representations through an \mlp, but these had no effect on performance.

This is an orthogonal way to incorporate context into the model relative to \pacrr. 
\pacrr creates a query-document similarity matrix and computes $n$-gram  convolutions over the matrix. Here we incorporate context directly into the term 
encodings; hence similarities in this space are already context-sensitive. One way to view this difference is the point at which context enters the model -- 
directly during term encoding (Eq.~\ref{eq:ct}) or after term similarity scores have been computed (\pacrr, Fig.~\ref{fig:pacrr}).


\subsection{ABEL-DRMM}
\label{sec:had-drmm}
\label{sec:abel-drmm}

Using the context-sensitive q-term and d-term encodings of $\S$\ref{sec:input-encodings} (Eq.~\ref{eq:ct}), our next extension to \drmm is to create \emph{document-aware} q-term encodings that go beyond bucketed histograms of cosine similarities, the stage in Fig.~\ref{fig:top-level-drmm} indicated by $\otimes$. We focus on differentiable encodings to facilitate end-to-end training from inputs to relevance scores.

\begin{figure}[t]
\includegraphics[width=3in]{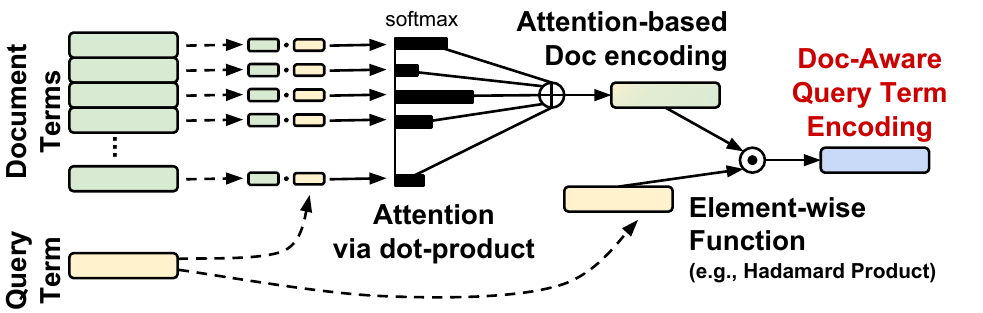}
\vspace{-0.12in}
\caption{\abeldrmm sub-net. From context-aware q-term and d-term encodings (Eq.~\ref{eq:ct}), it generates fixed-dimension \emph{document-aware q-term encodings} to be used in \drmm (Fig.~\ref{fig:top-level-drmm}, replacing $\otimes$ nodes).} 
\label{fig:hadamard-drmm}
\vspace{-4mm}
\end{figure}

Figure~\ref{fig:hadamard-drmm} shows the sub-network that computes the document-aware encoding of a q-term $q_i$ in the new model, given a document $d = \left<d_1, \ldots, d_m\right>$ of $m$ d-terms. We first compute a dot-product\footnote{Dot-products have a larger range than other similarity functions, encouraging low entropy attention distributions.} attention score $a_{i,j}$ for each $d_j$ relative to $q_i$:
\begin{equation}
a_{i,j} = \textrm{softmax}\Big(c(q_i)^T \, c(d_j); d_1, \dots, d_m\Big)
\label{eq:dpattention}
\end{equation}
where $c(t)$ is the context-sensitive encoding of $t$ (Eq.~\ref{eq:ct}). We then sum the context-sensitive encodings of the d-terms, weighted by their attention scores, to produce an attention-based representation $d_{q_i}$ of document $d$ from the viewpoint of $q_i$:
\begin{equation}
d_{q_i} = \sum_j a_{i,j} \; c(d_j)
\label{eq:dqi}
\end{equation}
The Hadamard product (element-wise multiplication, $\odot$) between 
the (L2-normalized) document representation $d_{q_i}$ and the q-term encoding $c(q_i)$ is then computed and used as the fixed-dimension document-aware encoding $\phi_{H}(q_i)$ of $q_i$ (Fig.~\ref{fig:hadamard-drmm}):
\begin{equation}
\phi_{H}(q_i) = \frac{d_{q_i}}{||d_{q_i}||} \odot \frac{c(q_i)}{||c(q_i)||}
\label{eq:hadamard}
\end{equation}
The $\otimes$ nodes and lower parts of the \drmm network of Fig.~\ref{fig:top-level-drmm} are now replaced by (multiple copies of) the sub-network of Fig.~\ref{fig:hadamard-drmm} (one copy per q-term), with the $\odot$ nodes replacing the $\otimes$ nodes. We call the resulting model Attention-Based ELement-wise \drmm (\abeldrmm).

Intuitively, if the document contains one or more terms $d_j$ that are similar to $q_i$, the attention mechanism will have emphasized mostly those terms and, hence, $d_{q_i}$ will be similar to $c(q_i)$, otherwise not. This similarity could have been measured by 
the cosine similarity between $d_{q_i}$ and $c(q_i)$, but the 
cosine similarity assigns the same weight to all the dimensions, i.e., to all the 
(L2 normalized) element-wise products in $\phi_{H}(q_i)$, which 
cosine similarity just sums. By using the Hadamard product, we pass on to the upper layers of \drmm (the dense layers of Fig.~\ref{fig:top-level-drmm}), which score each q-term with respect to the document, all the 
(normalized) element-wise products of $\phi_H(q_i)$, allowing the upper layers to learn which element-wise products (or combinations of them) are important when matching a q-term to the document.

Other element-wise functions can also be used to compare $d_{q_i}$ to $c(q_i)$, instead of the Hadamard product (Eq.~\ref{eq:hadamard}). For example, a vector containing the squared terms of the Euclidean distance between $d_{q_i}$ and $c(q_i)$ could be used instead of $\phi_H(q_i)$. This change had no effect on \abeldrmm's performance on development data. We also tried using $[d_{q_i};c(q_i)]$ instead of $\phi_{H}(q_i)$, but performance on development data deteriorated. 

\abeldrmm is agnostic to document length, like \drmm. 
\abeldrmm, however, is trainable end-to-end, unlike the original \drmm. 
Still, both models do not reward higher density matches. 


\subsection{POSIT-DRMM}
\label{sec:ps-drmm}
\label{sec:posit-drmm}

Ideally, we want models to reward both the maximum match between a q-term and a document, but also the average 
match (between several q-terms and the document) to reward documents that have a higher density of matches. The document-aware q-term scoring of \abeldrmm does not account for this, as the attention summation hides whether a single or multiple terms were matched with high similarity. We also want models to be 
end-to-end trainable, like \abeldrmm.

\begin{figure}[t]
\includegraphics[width=3in]{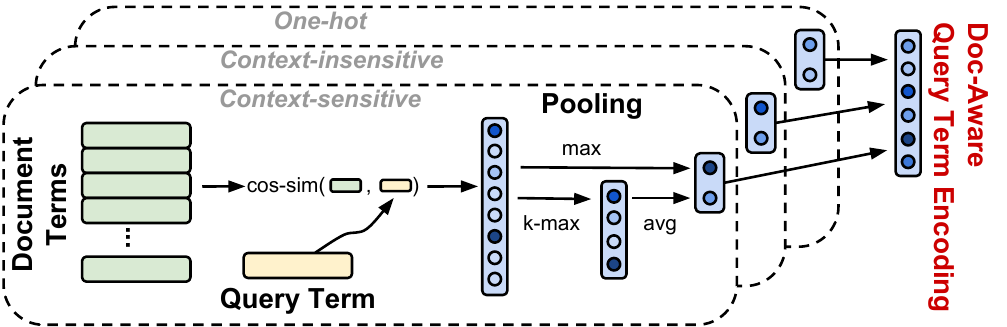}
\vspace{-0.15in}
\caption{\positdrmm with multiple views (\textsc{+mv}). Three two-dimensional document-aware q-term encodings, one from each view, are produced, concatenated, and used in \drmm (Fig.~\ref{fig:top-level-drmm}, replacing $\otimes$ nodes).}
\vspace{-4mm}
\label{fig:basic-ritr}
\end{figure}

Figure~\ref{fig:basic-ritr} (context-sensitive box) outlines a simple network that produces document-aware q-term encodings, 
replacing the \abeldrmm sub-network of Fig.~\ref{fig:hadamard-drmm} in the \drmm framework.
We call the resulting model POoled SImilariTy \drmm 
(\positdrmm). As in 
\abeldrmm, we compute an attention score $a_{i,j}$ for each $d_j$ relative to $q_i$, now using cosine similarity (cf.\ Eq.~\ref{eq:dpattention}):
\begin{equation}
a_{i,j} = 
\frac{c(q_i)^T c(d_j)}{||c(q_i)|| \; ||c(d_j)||}
\label{eq:cosattention}
\end{equation}
However, we do not use the $a_{i,j}$ scores to compute a weighted average of the encodings of the d-terms 
(cf.\ Eq.~\ref{eq:dqi}), which is also why there is no softmax in $a_{i,j}$ above (cf.\ Eq.~\ref{eq:dpattention}).\footnote{The $a_{i,j}$s still need to be normalized for input to the upper layers, but they do not need to be positive summing to 1. This is why we use cosine similarity in Eq.~\ref{eq:cosattention} instead of dot-products combined with softmax of Eq.~\ref{eq:dpattention}. } 
Instead, we concatenate the attention scores of the $m$ d-terms:
\[
a_i = \langle a_{i,1}, \dots, a_{i,j}, \dots, a_{i,m} \rangle^T
\]
and we apply two pooling steps on $a_i$ to create a 2-dimensional document-aware encoding $\phi_P(q_i)$ of the q-term $q_i$ (Fig.~\ref{fig:basic-ritr}). First max-pooling, which returns the single best match of $q_i$ in the document. Then average pooling over a $k$-max-pooled version of $a_i$, which represents the average similarity for the top $k$ matching terms: 
\[
\phi_P(q_i) = 
\Big\langle\mbox{max}(a_i), \textrm{avg}\Big(\mbox{k-max}(a_i)\Big)\Big\rangle^T
\]

\positdrmm has many fewer parameters than the other models. The input to the upper q-term scoring dense layers of the \drmm framwork (Fig.~\ref{fig:top-level-drmm}) for \abeldrmm has the same dimensionality as pre-trained term embeddings, on the order of hundreds. By contrast, the input dimensionality here is 2. 
Hence, \positdrmm does not require deep dense layers, but uses a single layer (depth 1).
More information on hyperparameters is provided in Appendix A (supplementary material).

\positdrmm is closely related to \pacrr (and \pacrrdrmm). Like \positdrmm, \pacrr first computes cosine similarities between all q-terms and d-terms (Fig.~\ref{fig:pacrr}). It then applies $n$-gram convolutions to the similarity matrix to inject context-awareness, and then pooling to create document-aware q-term representations. Instead, \positdrmm relies on the fact that the term encodings are now already context sensitive (Eq.~\ref{eq:ct}) and thus skips the $n$-gram convolutions. Again, this is a choice of when context is injected -- during term encoding or after computing similarity scores. 

Mohan et al.'s work (\citeyear{mohan2017deep}) is related in the sense that for each q-term, document-aware encodings are built over the best matching (Euclidean distance) d-term. But again, term encodings are context-insensitive pre-trained word embeddings and the model is not trained end-to-end.


\subsection{Multiple Views of Terms (+MV)}
\label{sec:multiview}

An extension to \abeldrmm and \positdrmm (or any deep model) is to use multiple views of terms. The basic \positdrmm produces a two-dimensional document-aware encoding of each q-term (Fig.~\ref{fig:basic-ritr}, context-sensitive box) viewing the terms as their context-sensitive encodings (Eq.~\ref{eq:ct}). Another two-dimensional document-aware q-term encoding can be produced by viewing the terms directly as their pre-trained  embeddings without converting them to context-sensitive encodings (Fig.~\ref{fig:basic-ritr}, context-insensitive box). A third view uses one-hot vector representations of terms, which allows exact term matches to be modeled, as opposed to near matches in embedding space.
Concatenating the outputs of the 3 views, we obtain 6-dimensional document-aware q-term encodings, leading to a model dubbed \positdrmmmv. An example of this multi-view document-aware query term representation is given in Fig.~\ref{fig:posit-example} for a query-document pair from \bioasq's development data.

The multi-view extension of \abeldrmm (\abeldrmmmv) is very similar, i.e., it uses context-sensitive term encodings, pre-trained term embeddings, and one-hot term encodings in its three views. The resulting three document-aware q-term embeddings can be summed or concatenated, though we found the former more effective. 


\subsection{Alternative Network Structures}

The new models 
($\S\S$\ref{sec:pacrr-drmm}--\ref{sec:multiview}) were selected by experimenting on development data. Many other extensions were considered, but not ultimately used as they were not beneficial empirically, including deeper and wider \rnn{s} or \cnn encoders \cite{bai2018empirical}; combining 
document-aware encodings from all models;  different attention mechanisms, e.g., multi-head \cite{vaswani2017attention}.

Pointer Networks \cite{vinyals2015pointer} use the attention 
scores directly to select an input component. 
\positdrmm does this via max and average pooling, not argmax.
We implemented Pointer Networks -- argmax over \abeldrmm attention to select the best d-term encoding -- but empirically this was similar to \abeldrmm.
Other architectures considered in the literature include the 
\textsc{k-nrm} model of \newcite{xiong2017end}. This 
is similar to both \abeldrmm and \positdrmm in that it can be viewed as an end-to-end version of \drmm. However, it uses kernels over the query-document interaction matrix to produce features per q-term.

The work of \newcite{pang2017deeprank} 
is highly related and investigates many different structures, specifically aimed at incorporating context-sensitivity. However, unlike our work, Pang et al.\ first extract contexts 
($n$-grams) of documents that match q-terms. Multiple interaction matrices are then constructed for the entire query relative to each of these contexts. These document contexts may match one or more q-terms allowing the model to incorporate term proximity. These interaction matrices can also be constructed using exact string match similar to \positdrmmmv.


\section{Experiments}

We experiment with ad-hoc retrieval datasets with hundreds of thousands or millions of documents.  
As deep learning models are computationally expensive, we first run a traditional \ir system\footnote{We used Galago (\url{http://www.lemurproject.org/galago.php}, v.3.10). We removed stop words and applied Krovetz's stemmer \cite{krovetz1993krovetzstemmer}.} using the \bmtf score \cite{robertson2009probabilistic} and then re-rank the top $N$ returned documents.

\subsection{Methods Compared}

All systems use an extension proposed by \newcite{severyn2015learning}, where the relevance score is combined via a linear model with a set of \emph{extra features}. We use four extra features: z-score normalized \bmtf score; percentage of 
q-terms with exact match in the document (regular and \idf weighted); and percentage of 
q-term bigrams matched in the document. The latter three features were taken from \newcite{mohan2017deep}.

In addition to the models  
of $\S\S$\ref{sec:drmm}, \ref{sec:pacrr}, \ref{sec:pacrr-drmm}--\ref{sec:multiview}, we used the following baselines:
Standard Okapi \bmtf (\textbf{BM25}); and \bmtf re-ranked with a linear model over the four extra features (\textbf{BM25+extra}).
These \ir baselines are very strong and most recently proposed deep learning models do \emph{not} beat them.\footnote{See, for example, Table~2 of \newcite{guo2016deep}.} 
\drmm and \pacrr are also strong baselines
and have shown superior performance over other deep 
learning models on a variety of data \cite{guo2016deep,hui2017pacrr}.\footnote{For \pacrr/\pacrrdrmm, we used/modified the code released by Hui et al.\ (\citeyear{hui2017pacrr}, \citeyear{hui2018copacrr}). We use our own implementation of \drmm, which performs roughly the same as \newcite{guo2016deep}, though the results are not directly comparable due to different random partitions of the data.}

All hyperparameters were tuned on 
development data and are available in Appendix A.
All models were trained using Adam \cite{kingma2014adam} with batches containing a randomly sampled negative example per positive example\footnote{We limit positive 
examples to be in the top $N$ documents.} 
and a pair-wise loss. 
As the datasets contain only documents marked as relevant, negative examples were sampled from the top $N$ documents (returned by \bmtf) that had not been marked as relevant.

We evaluated the models using the \trec ad-hoc retrieval evaluation script\footnote{\url{https://trec.nist.gov/trec\_eval/} (v9.0)} focusing on \map, Precision@20 and nDCG@20
\cite{manning2008ir}. We trained each model five times with different random seeds and report the mean and standard deviation for each metric on test data; in each run, the model selected had the highest \map on the development data.
We also report results for an \textbf{oracle}, which re-ranks the $N$ documents returned by \bmtf placing all human-annotated relevant documents at the top. 
To test 
for statistical significance between two systems, we employed two-tailed stratified shuffling 
\cite{smucker2007comparison,Dror2017HitchhikersGuide}
using the model with the highest development \map over the five runs per method.

\subsection{BioASQ Experiments}

Our first experiment used the dataset of the document ranking task of \bioasq \cite{tsatsaronis2015overview}, years 1--5.\footnote{\url{http://bioasq.org/}.} 
It contains 2,251 English biomedical questions,
each formulated by a biomedical expert, who  searched (via PubMed\footnote{\url{https://www.ncbi.nlm.nih.gov/pubmed/}}) for, and annotated relevant documents. Not all relevant documents were necessarily annotated, but the data includes additional expert relevance judgments made during the official evaluation.\footnote{Our results are, thus, not comparable to those of participating systems, since experts did not consider our outputs.}

\begin{table*}[t]
\footnotesize
\begin{minipage}{3.15in}
\begin{center}
\textbf{\textsc{{\normalsize average over five runs with std.\ dev.}}}\\
\vspace{0.2em}
\begin{tabular}{|llll|}
\hline
{\bf System}& {\bf MAP}    & {\bf P@20}   & {\bf nDCG@20} \\ \hline
\multicolumn{4}{|c|}{Traditional IR Baselines} \\ \hline
BM25        & $46.1\pm.0$ & $25.5\pm.0$ & $55.4\pm.0$ \\
BM25+extra  & $48.7\pm.0$ & $26.6\pm.0$ & $58.1\pm.0$ \\ \hline
\multicolumn{4}{|c|}{Deep Learning Baselines} \\ \hline
PACRR       & $49.1\pm.2$ & $27.1\pm.1$ & $58.5\pm.2$ \\
DRMM        & $49.3\pm.2$ & $27.2\pm.2$ & $58.5\pm.3$ \\ \hline
\multicolumn{4}{|c|}{Deep Learning with Enhanced Interactions} \\ \hline
PACRR-DRMM  & $49.9\pm.1$ & $27.4\pm.1$ & $59.3\pm.1$ \\
ABEL-DRMM   & $50.3\pm.2$ & $27.5\pm.1$ & $59.6\pm.2$ \\
$\quad$+MV  & $50.4\pm.2$ & $27.4\pm.2$ & $59.7\pm.3$\\
POSIT-DRMM  & $50.7\pm.2$ & $27.8\pm.1$ & $60.1\pm.2$ \\
$\quad$+MV  & $51.0\pm.1$ & $27.9\pm.1$ & $60.3\pm.2$ \\ \hline\hline
Oracle      & $72.8\pm.0$ & $37.5\pm.0$ & $80.7\pm.0$ \\ \hline
\end{tabular}
\end{center}
\end{minipage}
\begin{minipage}{3.1in}
\begin{center}
\textbf{\textsc{{\normalsize best run with stat.\ sig.}}}\\
\vspace{0.2em}
\begin{tabular}{|llll|}
\hline
{\bf System}& {\bf MAP}    & {\bf P@20}   & {\bf nDCG@20} \\ \hline
\multicolumn{4}{|c|}{Traditional IR Baselines} \\ \hline
BM25        & $46.1$ & $25.5$ & $55.4$ \\
BM25+extra  & $48.7\st{1}$ & $26.6\st{1}$ & $58.1\st{1}$ \\ \hline
\multicolumn{4}{|c|}{Deep Learning Baselines} \\ \hline
PACRR       & $49.1\st{1}$ & $27.0\st{1-2}$ & $58.6\st{1}$ \\
DRMM        & $49.3\st{1-2}$ & $27.1\st{1-2}$ & $58.8\st{1-2}$ \\ \hline
\multicolumn{4}{|c|}{Deep Learning with Enhanced Interactions} \\ \hline
PACRR-DRMM  & $50.0\st{1-3}$ & $27.3\st{1-2}$ & $59.4\st{1-3}$ \\
ABEL-DRMM   & $50.2\st{1-4}$ & $27.5\st{1-4}$ & $59.4\st{1-3}$ \\
$\quad$+MV  & $50.5\st{1-4}$ & $27.6\st{1-4}$ & $59.8\st{1-4}$ \\
POSIT-DRMM  & $50.7\st{1-4,6}$ & $27.9\st{1-7}$ & $60.1\st{1-4,6}$ \\
$\quad$+MV  & $51.0\st{1-7}$ & $27.7\st{1-4}$ & $60.3\st{1-7}$\\ \hline\hline
Oracle      & $72.8$ & $37.5$ & $80.7$ \\ \hline
\end{tabular}
\end{center}
\end{minipage}
\vspace{-0.1in}
\caption{Performance on \bioasq test data. Statistically significant ($p < 0.05$)
difference from  \bmtf$\!\!\!^1$; {\sc bm25}+extra$^2$; \pacrr$\!\!^3$; \drmm$\!^4$; \pacrrdrmm$\!^5$; \abeldrmm$\!^6$; \abeldrmmmv$\!^7$.}
\vspace{-1mm}
\label{tab:bioasq-main}
\end{table*}

\begin{table*}[t]
\footnotesize
\begin{minipage}{3.15in}
\begin{center}
\textbf{\textsc{{\normalsize average over five runs with std.\ dev.}}}\\
\vspace{0.2em}
\begin{tabular}{|llll|}
\hline
{\bf System}& {\bf MAP}    & {\bf P@20}   & {\bf nDCG@20} \\ \hline
\multicolumn{4}{|c|}{Traditional IR Baselines} \\ \hline
BM25        & $23.8\pm.0$ & $35.4\pm.0$ & $42.5\pm.0$ \\
BM25+extra  & $25.0\pm.0$ & $36.7\pm.0$ & $43.2\pm.0$ \\ \hline
\multicolumn{4}{|c|}{Deep Learning Baselines} \\ \hline
PACRR       & $25.8\pm.2$ & $37.2\pm.4$ & $44.3\pm.4$ \\
DRMM        & $25.6\pm.6$ & $37.0\pm.8$ & $44.4\pm.6$\\ \hline
\multicolumn{4}{|c|}{Deep Learning with Enhanced Interactions} \\ \hline
PACRR-DRMM  & $25.9\pm.4$ & $37.3\pm.7$ & $44.4\pm.7$ \\
ABEL-DRMM   & $26.3\pm.4$ & $38.0\pm.6$ & $45.6\pm.4$ \\
$\quad$+MV  & $26.5\pm.4$ & $38.0\pm.5$ & $45.5\pm.4$ \\
POSIT-DRMM  & $27.0\pm.4$ & $38.3\pm.6$ & $45.7\pm.5$\\
$\quad$+MV  & $27.2\pm.3$ & $38.6\pm.6$ & $46.1\pm.4$\\ \hline\hline
Oracle      & $68.0\pm.0$ & $82.1\pm.0$ & $93.1\pm.0$ \\ \hline
\end{tabular}
\end{center}
\end{minipage}
\begin{minipage}{3.1in}
\begin{center}
\textbf{\textsc{{\normalsize best run with stat.\ sig.}}}\\
\vspace{0.2em}
\begin{tabular}{|llll|}
\hline
{\bf System}& {\bf MAP}    & {\bf P@20}   & {\bf nDCG@20} \\ \hline
\multicolumn{4}{|c|}{Traditional IR Baselines} \\ \hline
BM25        & $23.8$ & $35.4$ & $42.5$ \\
BM25+extra  & $25.0\st{1}$ & $36.7\st{1}$ & $43.2\st{1}$ \\ \hline
\multicolumn{4}{|c|}{Deep Learning Baselines} \\ \hline
PACRR       & $25.8\st{1-2}$ & $37.4\st{1-2}$ & $44.5\st{1-2}$ \\
DRMM        & $25.9\st{1-2}$ & $37.2\st{1-2}$ & $44.4\st{1-2}$\\ \hline
\multicolumn{4}{|c|}{Deep Learning with Enhanced Interactions} \\ \hline
PACRR-DRMM  & $25.9\st{1-2}$ & $37.6\st{1-2,4}$ & $44.5\st{1-2}$ \\
ABEL-DRMM   & $26.1\st{1-5}$ & $38.0\st{1-5}$ & $45.4\st{1-5}$ \\
$\quad$+MV  & $26.4\st{1-6}$ & $38.2\st{1-5}$ & $45.8\st{1-6}$ \\
POSIT-DRMM  & $27.1\st{1-7}$ & $38.8\st{1-7}$ & $46.2\st{1-7}$\\
$\quad$+MV  & $27.1\st{1-7}$ & $38.9\st{1-7}$ & $46.4\st{1-7}$\\ \hline\hline
Oracle      & $68.0$ & $82.1$ & $93.1$ \\ \hline
\end{tabular}
\end{center}
\end{minipage}
\vspace{-0.1in}
\caption{Performance on \trecrob 
test data. Statistically significant ($p < 0.05$) 
difference from  \bmtf$\!\!^1$; {\sc bm25}+extra$^2$; \pacrr$\!\!^3$; \drmm$\!^4$; \pacrrdrmm$\!^5$; \abeldrmm$\!^6$; \abeldrmmmv$\!^7$.}
\label{tab:rob04-test}
\vspace{-2mm}
\end{table*}

The document collection consists of approx.\
28M `articles' (titles and abstracts only) from the `\medline/PubMed Baseline 2018' collection.\footnote{See \url{https://www.nlm.nih.gov/databases/download/pubmed_medline.html}.} We discarded the approx.\ 10M articles that contained only titles, since very few of these were annotated as relevant. For  the remaining 18M articles, a document was the concatenation of each title and abstract. Consult Appendix B for further statistics of the dataset. Word embeddings were pre-trained by applying word2vec \cite{mikolov2013distributed} (see Appendix A for hyper-parameters) to the 28M `articles' of the \medline/PubMed collection. \idf values were computed over the 18M articles that contained both titles and abstracts.

The 1,751 queries of years 1--4 were used for training, the first 100 queries of year 5 (batch 1) for development, and the remaining 400 queries of year 5 (batches 2--5) as test set. We set $N=100$, since even using only the top 100 documents of \bmtf, the oracle scores are high. 
PubMed articles published after 2015 for the training set, and after 2016 for the development and test sets, were removed from the top $N$ (and replaced by lower ranked documents up to $N$), as these were not available at the time of the human annotation.

Table~\ref{tab:bioasq-main} reports results on the \bioasq test set, averaged over five runs as well as the single best run (by development \map) with statistical significance. 
The enhanced models of this paper perform better than \bmtf (even with extra features), \pacrr, and \drmm. There is hardly any difference between \pacrr and \drmm, but our combination of the two (\pacrrdrmm) surpasses them both on average, 
though the difference is statistically significant ($p < 0.05$) only when comparing to \pacrr. 
Models that use context-sensitive term encodings (\abeldrmm, \positdrmm) outperform other models, even \pacrr-style models that incorporate context at later stages in the network. This is true both on average and by statistical significance over the best run. The best model on average is \positdrmmmv, though it is not significantly different than \positdrmm.

\subsection{TREC Robust 2004 Experiments}

Our primary experiments were on the \bioasq 
dataset as it has one of the largest 
sets of queries (with manually constructed relevance judgments) and document collections, making it 
a particularly realistic dataset. However, in order to ground our models in past work we also ran experiments on \trecrob 2004 \cite{voorhees2005trec}, which is a common benchmark. It contains 250 
queries\footnote{We used the `title' fields of the queries.}
and 528K documents. As this dataset is quite small, we used a 5-fold cross-validation. In each fold, approx.\ $\frac{3}{5}$ of the queries were used for training, $\frac{1}{5}$ for development, and $\frac{1}{5}$  for testing. We applied word2vec to the 528K documents to obtain pre-trained embeddings. \idf values were computed over the same corpus. Here we used $N=1000$, as the oracle scores for $N=100$ were low.

Table~\ref{tab:rob04-test} shows the \trecrob results, which largely mirror those of \bioasq. \positdrmmmv is still the best model, though again not significantly 
different than \positdrmm. Furthermore, \abeldrmm and \positdrmm are clearly better than the deep learning baselines,\footnote{The results we report for our implementation of \drmm are slightly different than 
those of \newcite{guo2016deep}. There are a number of reasons for why this might be the case: there is no standard split of the data; non-standard preprocessing of the documents; the original \drmm paper reranks the top documents returned by Query Likelihood and not BM25.} 
but unlike \bioasq, 
there is no statistically significant difference between \pacrrdrmm and the two deep learning baselines.
Even though the scores are quite close (particularly \map) both \abeldrmm and \positdrmm are statistically different from \pacrrdrmm, which was not the case for \bioasq.
\abeldrmmmv is significantly 
different than \abeldrmm on the best run for \map and nDCG@20, unlike \bioasq where there was no 
statistically significant difference between the 
two methods. However, on average over 5 runs, the systems show little difference.


\section{Discussion}

\begin{figure*}[th]
\begin{minipage}{4.1in}
\includegraphics[width=4in]{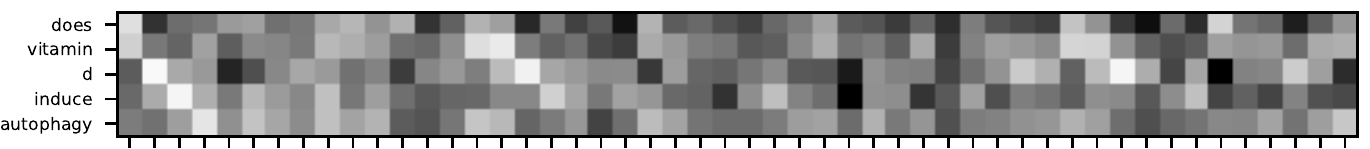}
\includegraphics[width=4in]{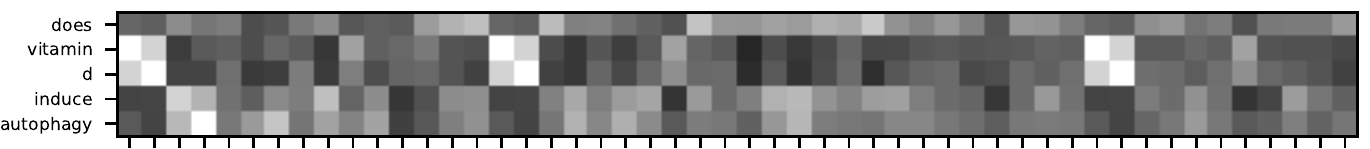}
\includegraphics[width=4in]{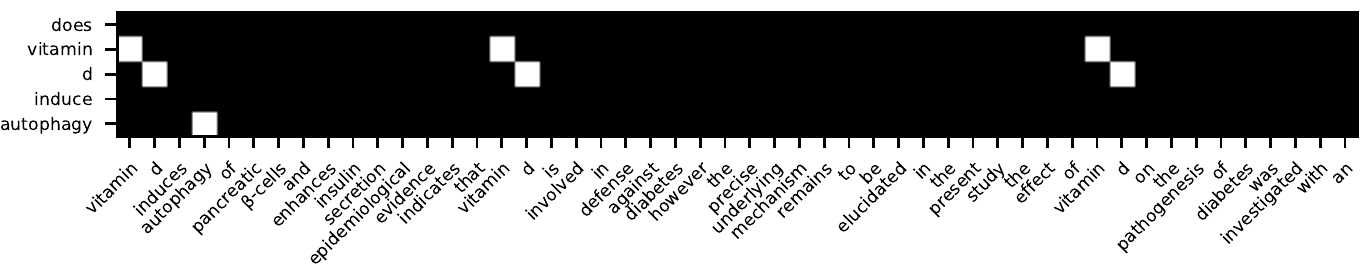}
\end{minipage}
\begin{minipage}{2in}
\includegraphics[width=1.9in]{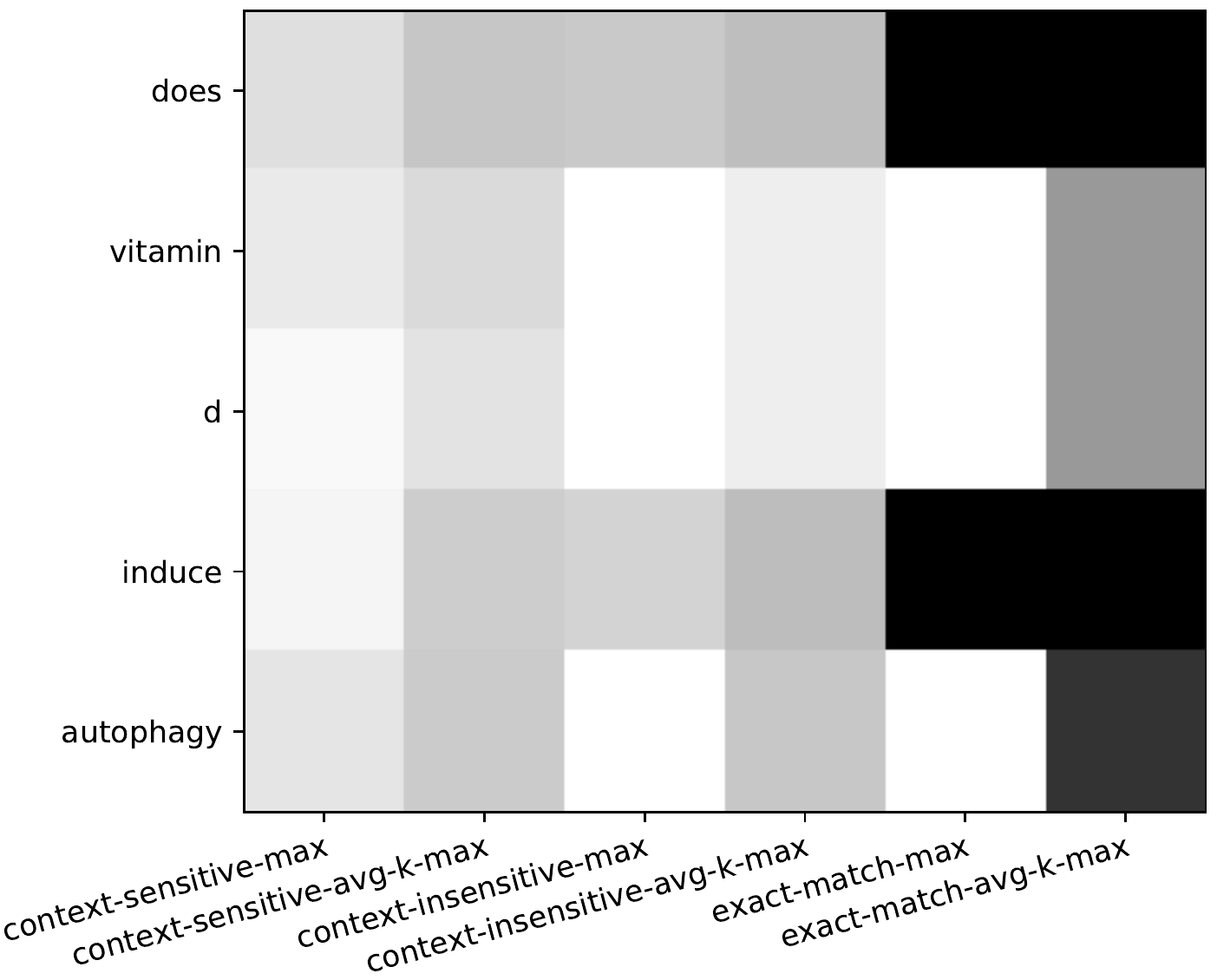}
\vspace{0.1in}
\end{minipage}
\vspace{-0.1in}
\caption{Left: Cosine similarities (\positdrmm attention) of query and document terms, with context-sensitive, context-insensitive, and exact match views of the terms (top to bottom). Document truncated to 50 words. White is stronger. Right: Corresponding \positdrmmmv 6-dimensional document-aware query term encodings.}
\label{fig:posit-example}
\vspace{-0.1in}
\end{figure*}

\begin{figure}[t]
\begin{center}
\includegraphics[width=1.9in]{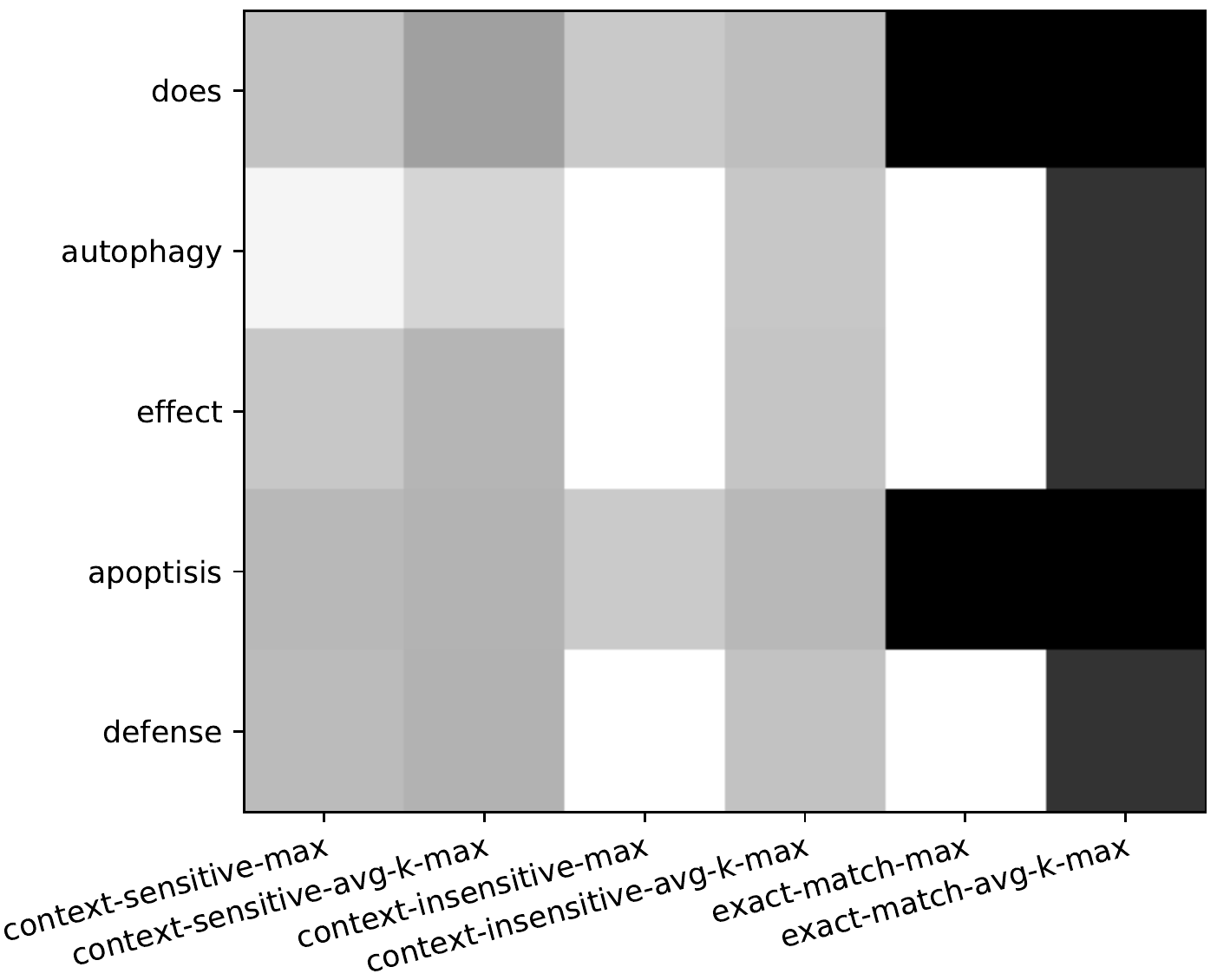}
\end{center}
\vspace{-0.2in}
\caption{\positdrmmmv 6-dimensional document-aware q-term encodings for  `Does autophagy induce apoptosis defense?' and the same document as Fig.~\ref{fig:posit-example}.}
\label{fig:posit-bad-example}
\vspace{-0.2in}
\end{figure}

An interesting question is how well the deep models do without the extra features. For \bioasq, the best model's (\positdrmmmv) \map score drops from 48.1 to 46.2 on the development set, which is higher than the \bmtf baseline (43.7), but on-par with \textsc{bm25+extra} (46.0).
We should note, however, that on this set, the \drmm baseline without the extra features (which include \bmtf) is actually lower than \bmtf (\map 42.5), though it is obviously adding a useful signal, since \drmm with the extra features performs better (46.5).

We also tested the contribution of context-sensitive term encodings (Eq.~\ref{eq:ct}). Without them, i.e., using directly the pre-trained embeddings, \map on \bioasq development data dropped from 47.6 to 46.3, and from 48.1 to 47.0 for \abeldrmm and \positdrmm, respectively.

Fig.~\ref{fig:posit-example} shows the cosine similarities (attention scores, Eq.~\ref{eq:cosattention}) between q-terms and d-terms, using term encodings of the three views (Fig.~\ref{fig:basic-ritr}), for a query ``Does Vitamin D induce 
autophagy?'' and a relevant document from the \bioasq development data. \positdrmm indeed marks this as relevant. In the similarities of the context-insensitive view (middle left box) we see multiple matches around `vitamin d' and `induce autophagy'. The former is an exact match (white squares in lower left box) and the latter a soft match. The context-sensitive view (upper left box) smooths things out and one can see a straight diagonal white line matching `vitamin d induce autophagy'. 
The right box of Fig.~\ref{fig:posit-example} shows the 6 components (Fig.~\ref{fig:basic-ritr}) of the document-aware q-term encodings. Although some terms are not matched exactly, the context sensitive max and average pooled components (two left-most columns) are high for all q-terms. Interestingly, `induce' and `induces' are not an exact match (leading to black cells for `induce' in the two right-most columns) and the corresponding context-insensitive component of (third cell from left) is low. However, the two components of the context-sensitive view (two left-most cells of `induce') are high, esp.\ the max-pooling component (left-most).Finally, `vitamin d' has multiple matches leading to a high average k-max pooled value, which indicates that the importance of that phrase in the document.  

Fig.~\ref{fig:posit-bad-example} shows the 6 components of the document-aware q-term encodings for 
another query and the same document, which is now irrelevant. In the max pooling columns of the exact match and context-insensitive view (columns 3, 5), the values look quite similar to those of Fig.~\ref{fig:posit-example}. However, \positdrmm scores this query-document pair low for two reasons. First, in the average-k-max pooling columns (columns 2, 4, 6) we get 
lower values than Fig.~\ref{fig:posit-example}, indicating that there is less support for this pair in terms of density. Second, the context sensitive values (columns 1, 2) are much worse, indicating that even though many exact matches exist, in context, the meaning is not the same.

We conclude by noting there is still quite a large gap between the current best models and the oracle re-ranking scores. Thus, there is head room for improvements through more data or better models.

\section*{Acknowledgements}
\vspace{-0.1in}
We thank the reviewers for their constructive feedback that greatly improved this work. Oscar T\"{a}ckstr\"{o}m gave thorough input on an early draft of this work. Finally, \textsc{aueb}'s \textsc{nlp} group provided many suggestions over the course of the work.


\bibliography{emnlp2018}
\bibliographystyle{acl_natbib_nourl}
\end{document}